\documentclass[%
 preprint,
superscriptaddress,
groupedaddress,
unsortedaddress,
runinaddress,
preprint,
 amsmath,amssymb,
 aps,
]{revtex4}
\usepackage{mathtools}
 
\usepackage{graphicx}
\usepackage{multirow}
\usepackage{titlesec}
\usepackage{bm}

\usepackage{hyperref}
\hypersetup{colorlinks,linkcolor={blue},citecolor={blue},urlcolor={blue}} 
\usepackage[labelfont=bf]{caption}

\begin{document}

\title{Real and Imaginary Phase Shifts for Nucleon-Deuteron Scattering using Phase Function Method}
\author{Shikha Awasthi and O. S. K. S. Sastri}
 \altaffiliation[E-mail: ]{sastri.osks@hpcu.ac.in}
\affiliation{Department of Physics and Astronomical Sciences, Central University of Himachal Pradesh, Dharamshala, 176215,H.P., Bharat India
}%


\begin{abstract}
{\bf{Abstract\textendash}} The neutron-deuteron ($nd$) and proton-deuteron ($pd$) scattering are the simplest nucleon-nucleus scenario which throws light on understanding few body systems. In this work, real and imaginary parts of scattering phase shifts (SPS) for $nd$ and $pd$ scattering are obtained using optical potential, with Malfliet-Tjon (MT) model of interaction, by phase function method (PFM). The SPS for doublet $^2S_{1/2}$ and quartet $^4S_{3/2}$ states of \textit{nd} and \textit{pd} systems have been obtained for real and imaginary parts separately by solving the phase equation for $\ell=0$, using Runge-Kutta $5^{th}$ order technique for laboratory energies upto 19 $MeV$. The obtained (real, imaginary) SPS for $^2S_{1/2}$ and $^4S_{3/2}$ states are matching with standard data with mean absolute error (MAE) of (1.32, 0.06) for $^2S_{1/2}$ state and (0.19, 0.06) for $^4S_{3/2}$ state of $nd$ scattering, (1.47, 0.62) for $^2S_{1/2}$ state and (0.55, 0.14) for $^4S_{3/2}$ state of $pd$ scattering 
\end{abstract}
\maketitle
\section{Introduction}
One of the most crucial methods for determining the nature of interactions between various particles or systems of particles, which in turn would give information about their internal structure, is scattering experiments. The scattering phase shifts (SPS) obtained from phase wave analysis carry the specifics of the interaction for various $\ell$-channels and are used to determine the differential and total scattering cross-sections. Quantum mechanically, the interaction potential between two particles is modeled and corresponding time independent Schrodinger equation (TISE) is solved to obtain the wavefunction. The obtained wavefunction is matched with asymptotic free particle wavefunction, to determine SPS. Correct match between the obtained and experimental phase shifts validates the model of interaction potential.\\
Study of neucleon-deuteron scattering is one of the extensive fields of research and many research groups have studied it both theoretically and experimentally \cite{Payne, Chen, Kievsky, Golak, Ishikawa} over the years. SPS have been obtained  for $nd$ system using different phenomenological potentials such as Manning-Rosen \cite{Khirali} and Hulthen \cite{Hulthen}. Alternatively, Fadeev equations have been solved for the three body-problem using AV18 nucleon-nucleon potential \cite{ZM} and Malfliet-Tjon \cite{MT} potential \cite{MT, Ishikawa, Payne}. Typically, all these studies focus on below deuteron break up threshold of 3 $MeV$ \cite{Chen, Ishikawa, Payne, Kievsky}. Chen $et.$ $al$ \cite{ZM} performed phase shifts analysis for proton-deuteron for proton lab energies up to 22.7 $MeV$ using AV18 nucleon-nucleon potential. Pisa group \cite{Viviani} has calculated proton-deuteron scattering amplitudes for energies of 5 and 10 $MeV$ using Kohn variational principle. Huber $et.el.$ \cite{Huber} and J. Arvieux \cite{Arvieux} did extensive phase shifts analysis for neutron-deuteron ($nd$) and proton-deuteron ($pd$) systems for both below and above three-body break up threshold in the elastic region, and their data is generally recognized as a standard data.\\
For scattering above the breakup threshold, imaginary part of SPS contribute significantly to total SCS and hence they can not be neglected. Since, above or even very close to the threshold, the imaginary part has a very rapid growth \cite{Huber}. The concept of optical potential for understanding nucleon-nucleon interactions has been introduced by many researchers \cite{Darriulat, Jana, Kuros}. Darriulat $et.al.$ \cite{Darriulat} and A.K. Jana \cite{Jana} have obtained imaginary SPS for $\alpha-\alpha$ scattering by considering optical potential with Woods-Saxon potential as the interaction potential.\\
We have previously obtained real part of SPS for doublet state of $nd$ and $pd$ systems for lab energies ranging between (1-10) $MeV$ using molecular Morse potential as model of interaction \cite{Triton, pd}. In present work, considering Malfliet-Tjon potential (MT) \cite{MT} as model of interaction, for both parts of optical potential, we obatin SPS for both doublet and quartet states of $nd$ and $pd$ scattering for laboratory energies  ranging from 1 to 19 $MeV$ using phase function method (PFM) \cite{Calogero, Babikov}. \\
\section{Methodology}
\subsection{Modeling the interaction}
Considering the nucleon-deuteron systems to be simplified to a one body system, the phenomonological interaction potential would be consisting of nuclear and Coulomb contributions. Since, we are considering energies beyond deuteron threshold, one has to determine real and imaginary SPS separately, by choosing appropriate mathematical functions for real and imaginary parts of an optical potential. To study neutron-deuteron (\textit{nd}) interaction below 3 $MeV$ threshold, where only real part of SPS exists, the interaction between neutron and deuteron is modeled using Malfliet-Tjon (MT) potential, given by:
\begin{equation}
V(r)= V_{MT}(r)= -V_A\Big(\frac{e^{-\mu_A r}}{r}\Big)+ V_R\Big(\frac{e^{-\mu_R r}}{r}\Big)
\label{MT}
\end{equation}
where, $\mu_R = 2\mu_A$ in units of $fm^{-1}$ and it has given reasonably good results\cite{MT, DAE}. So, the same has been considered as model of interaction for both real and imaginary parts of optical potential, in this work:
\begin{equation}
V_N(r) = V_R(r) + i V_I(r)
\end{equation}
For, nd system, it is given by
\begin{equation}
V_{nd}(r) =  -V_{AR}\Big(\frac{e^{-\mu_{AR} r}}{r}\Big)+ V_{RR}\Big(\frac{e^{-\mu_{RR} r}}{r}\Big) - i \Bigg( -V_{AI}\Big(\frac{e^{-\mu_{AI} r}}{r}\Big)+ V_{RI}\Big(\frac{e^{-\mu_{RI} r}}{r}\Big)\Bigg)
\label{Opticalpot}
\end{equation}
For proton-deuteron (\textit{pd}) interaction in addition to the nuclear interaction, one needs to account for Coulomb interaction due to protons. We consider an ansatz for Coulomb interaction as:
\begin{equation}
V_C(r) = z_1 z_2 \frac{e^2}{r} erf(\beta r)
\end{equation} 
which has been utilised for $\alpha-\alpha$ scattering by Buck $et.al.$ \cite{Buck}, Ali Bodmer $et.al.$ \cite{Ali} and Anil $et.al.$ \cite{alpha} with reasonable success. Here for \textit{pd} system, we have $z_1 = z_2 = 1$, $e^2=1.44$ $MeV fm$ and $\beta = \frac{\sqrt{3}}{2 \times R_{pD}}$, with $R_{pd}=1.9642 fm$ \cite{Morton} being its root mean square (rms) radius, resulting in $\beta = 0.441$ $fm^{-1}$. The erf() function has the characteristic of taking the shape of Coulomb potential after distance $R_{pd}$. To cut short the long range of Coulomb potential to a certain distance, say $r_f$, beyond which it is negligible compared to nuclear interaction, one can apply the condition \cite{Okai} that $\left|\frac{V_N(r_f)}{V_C(r_f)}\right|$ ratio is less than or equal to $10^{-5}$, where $V_N$ is the nuclear interaction potential taken from $nd$ system.
The total interaction potential for proton-deuteron (\textit{pd}) system is thus:
 \begin{equation}
        V_{pd}(r)= V_N(r)+ V_C(r)
    \end{equation}
\subsection{Phase Function Method} 
The radial TISE for $\ell = 0$, is given by
\begin{equation}
\frac{d^2u_0(r)}{dr^2} + \frac{2\mu}{\hbar^2}(E-V(r))u_0(r) = 0
\end{equation} 
Here, it is required that $u_0(r=0)=0$.
The free particle asymptotic wavefunction for $r > R$, when V(r) becomes zero, would be 
\begin{equation}
u_a(r) = C sin(kr + \delta)
\label{asym}
\end{equation}
where constants $C$ and $\delta$ are to be obtained using boundary conditions. Here, $k$ is related to centre-of-mass energy by relation: 
\begin{equation}
k = \sqrt{\frac{E_{cm}}{\hbar^2 / 2 \mu}}
\end{equation}
 and
\begin{equation}
E_{cm}= \Big(\frac{m_T}{m_P+m_T}\Big)E_{lab} 
\end{equation}
Here, $m_P$ and $m_T$ are the masses of projectile and target respectively.

Let us consider the derivative of logarthmic wavefunction 
\begin{equation}
\frac{1}{u_0}\frac{du_0}{dr} = \frac{1}{u_a}\frac{du_a}{dr} = A(r)
\label{impedance}
\end{equation}
to be continuous at the boudnary. This is equivalent to impedance matching \cite{impedance} across the boundary.
Considering the first term, one obtains derivative of A(r) to be 
\begin{equation}
\frac{dA(r)}{dr} = -A^2(r) + \frac{1}{u_0}\frac{d^2u_0(r)}{dr^2}
\end{equation} 
Substituting this in TISE, we obtain
\begin{equation}
\frac{dA(r)}{dr} + A^2(r) = \frac{2\mu}{\hbar^2}(V(r)-E)
\label{tise1}
\end{equation}

Now, considering second term in Eq. \ref{impedance}, one obtains
\begin{equation}
A(r) = k \cot(kr + \delta)
\label{Ar}
\end{equation}
and 
\begin{equation}
\frac{dA(r)}{dr} = -\frac{(k^2 + k\frac{d\delta}{dr})}{\sin^2(kr + \delta)}
\label{Ader}
\end{equation}
Substituting Eqs. \ref{Ar} and \ref{Ader} into Eq. \ref{tise1}, one obtains phase equation as  
\begin{equation}
\delta'(k,r) = -\frac{U(r)}{k} \sin^2(kr +\delta(k,r))
\label{delta}
\end{equation}
where $U(r) = \frac{2\mu V(r)}{\hbar^2}$.\\
This is the final equation to be solved numerically from origin to asymptotic region where the effect of potential on scattering phase shifts is negligible, by using RK-5 method \cite{rk5} in order to obtain scattering phase shifts for S-wave.
\subsection{Scattering parameters}
The Scattering parameters i.e scattering length `$a$' and effective range `$r_0$' are calculated by using relation \cite{Darewych}
\begin{equation}
k \cot(\delta) = -\frac{1}{a} + 0.5 r_0 k^2
\label{reg}
\end{equation}
A plot of $0.5 k^2$ $vs$ $k\cot(\delta)$ would allow calculation of $r_0$ and $a$ from its slope and intercept respectively.
\section{RESULTS AND DISCUSSION}
\begin{figure*}[h]
\centering
{\includegraphics[height = 7 cm, width =8.1cm]{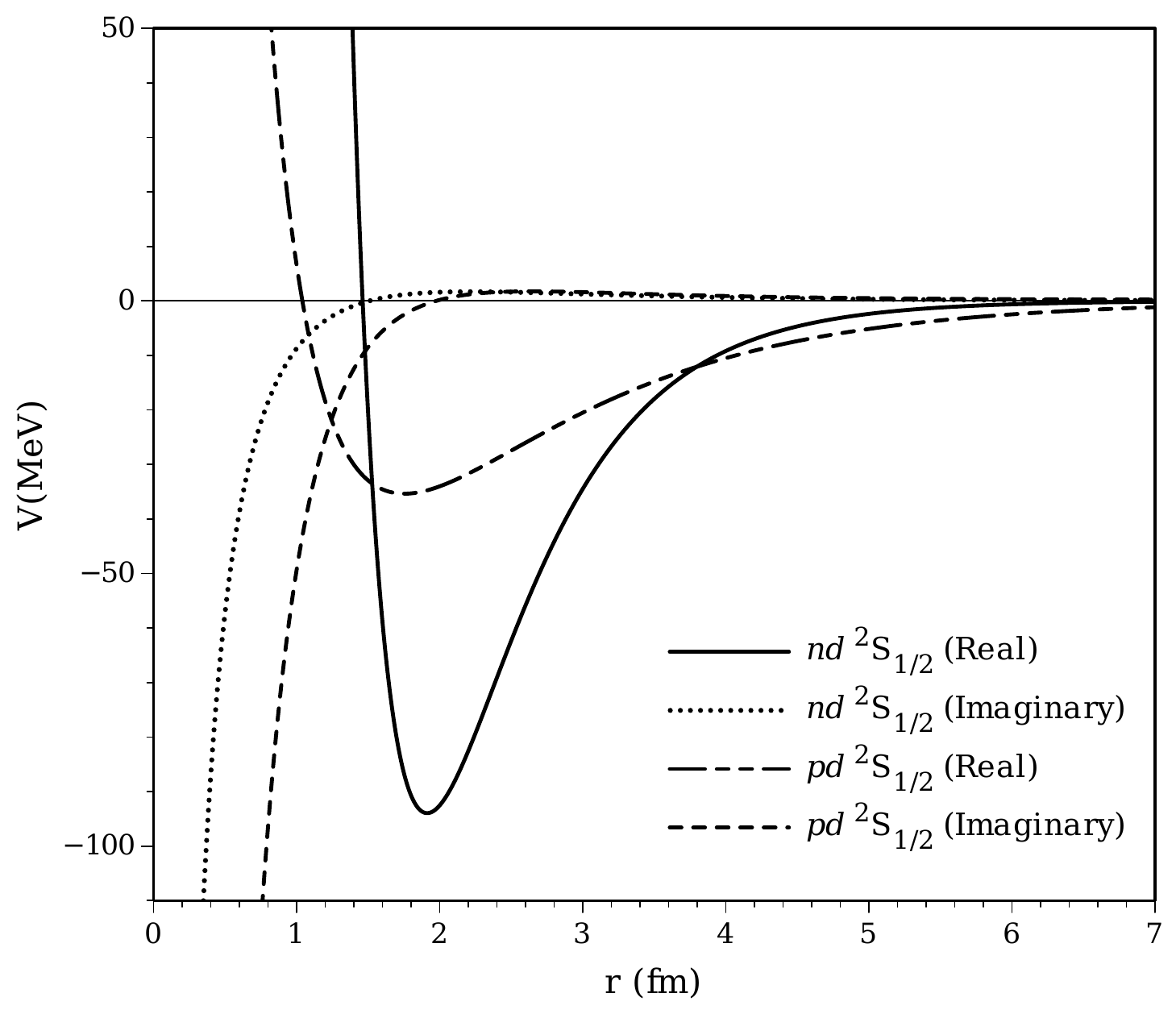}  }
{\includegraphics[height = 7 cm, width = 8.1cm]{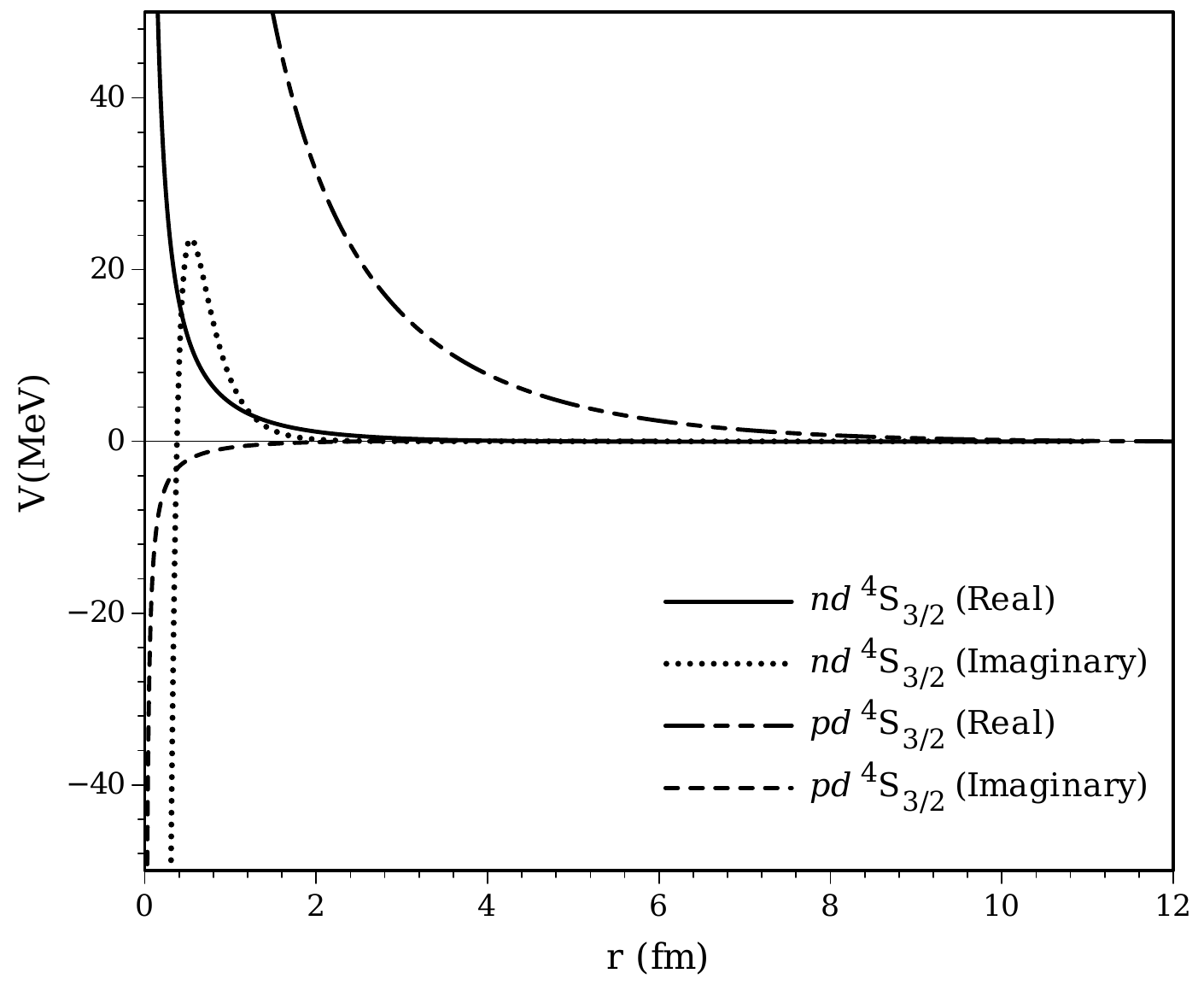}  }
\caption{Real and imaginary potentials for $^2S_{1/2}$ and $^4S_{3/2}$ states of $nd$ and $pd$ systems, w.r.t interaction distance \textit{r} by using optimised parameters given in Table \ref{parameters}.}
\label{fig1}
\end{figure*}

\begin{figure*}[htp]
\centering
{\includegraphics[height = 7 cm, width =8.1cm]{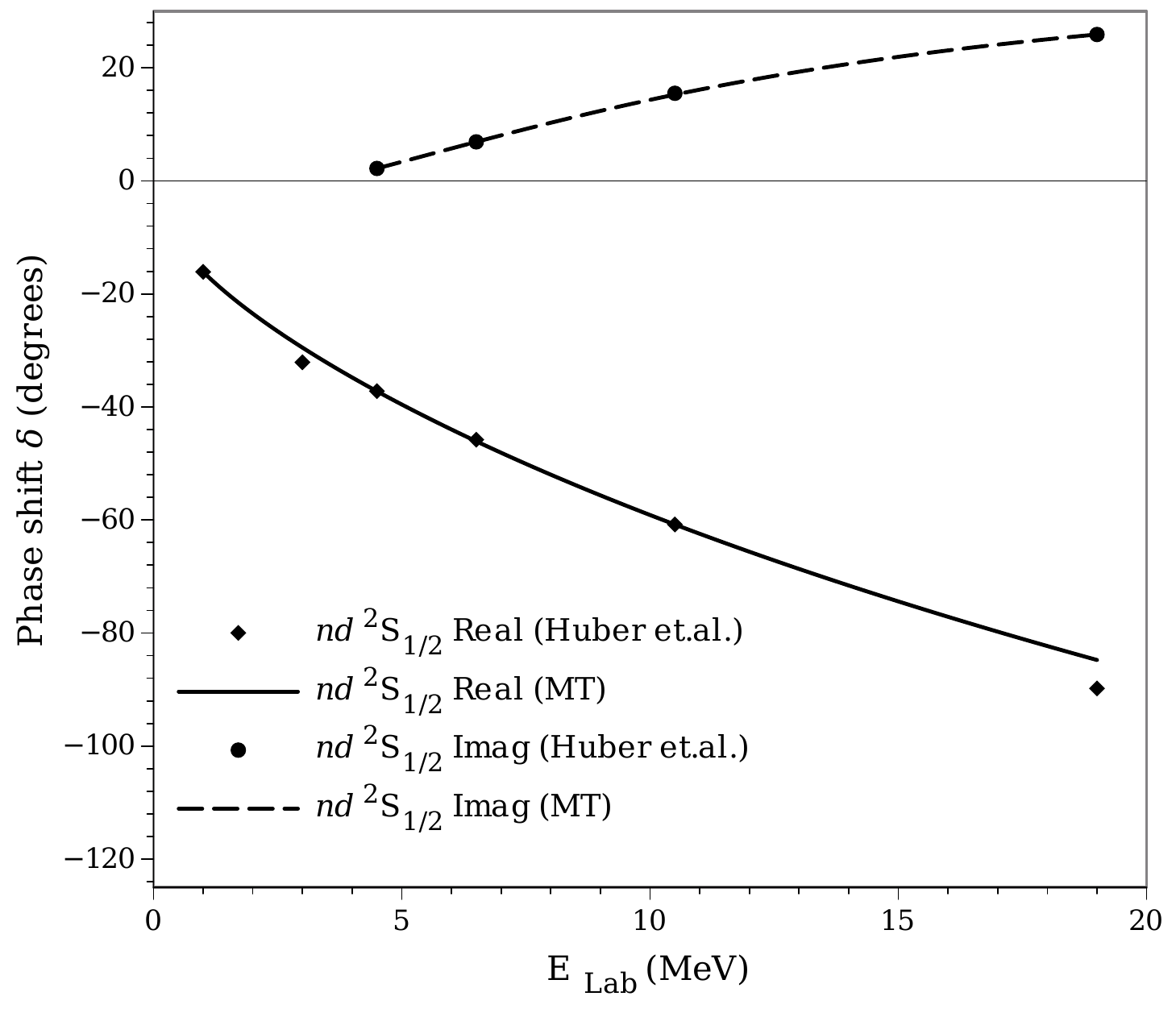}  }
{\includegraphics[height = 7 cm, width = 8.1cm]{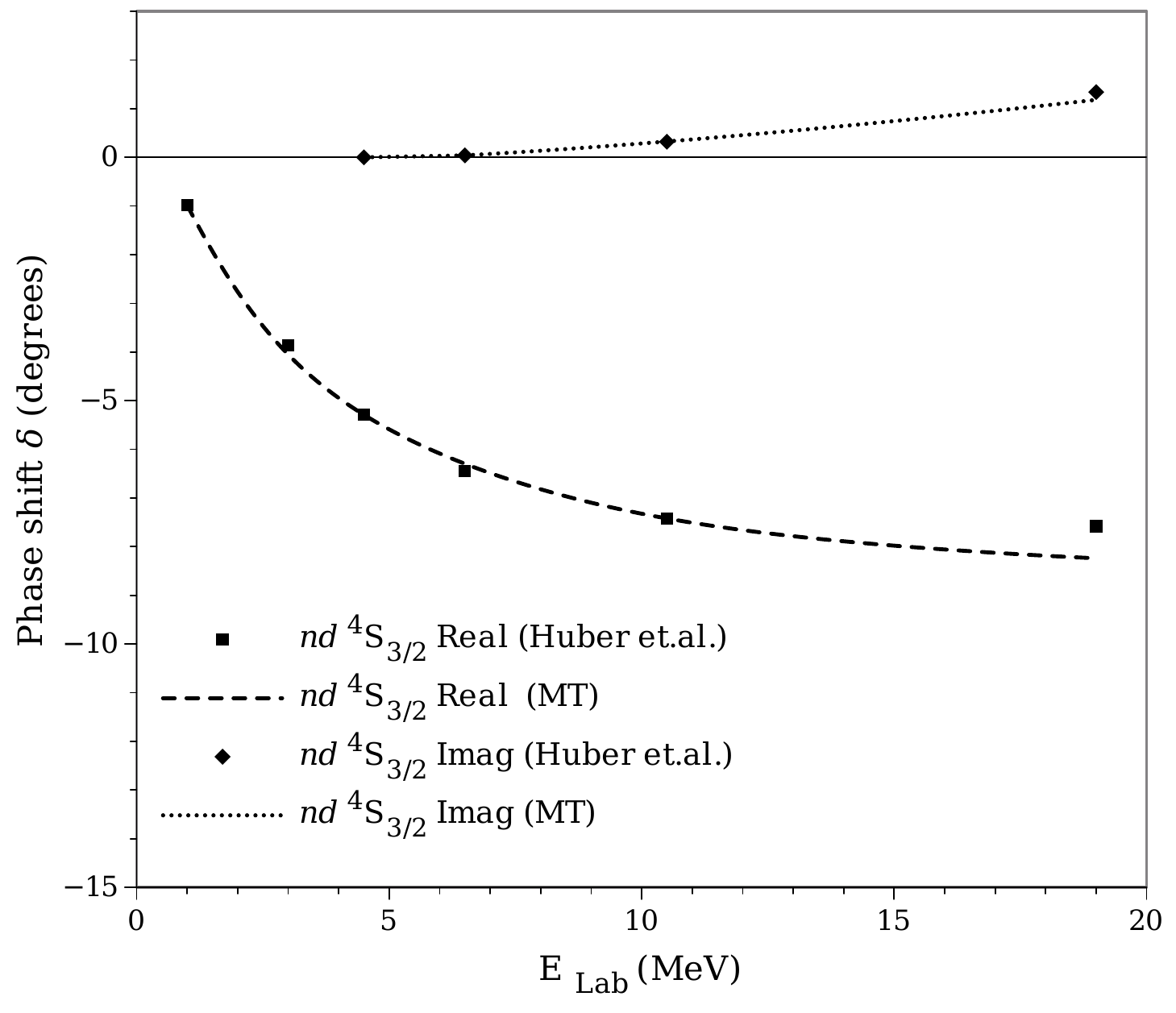}  }
\caption{Real and imaginary parts of scattering phase shifts (SPS) for $^2S_{1/2}$ (\textit{left}) and $^4S_{3/2}$ (\textit{right}) states of $nd$ system, plotted against laboratory energy ranging between (1-19) $MeV$.}
\label{fig3}
\end{figure*}

\begin{figure*}[htp]
\centering
{\includegraphics[height = 7.12 cm, width =8.1cm]{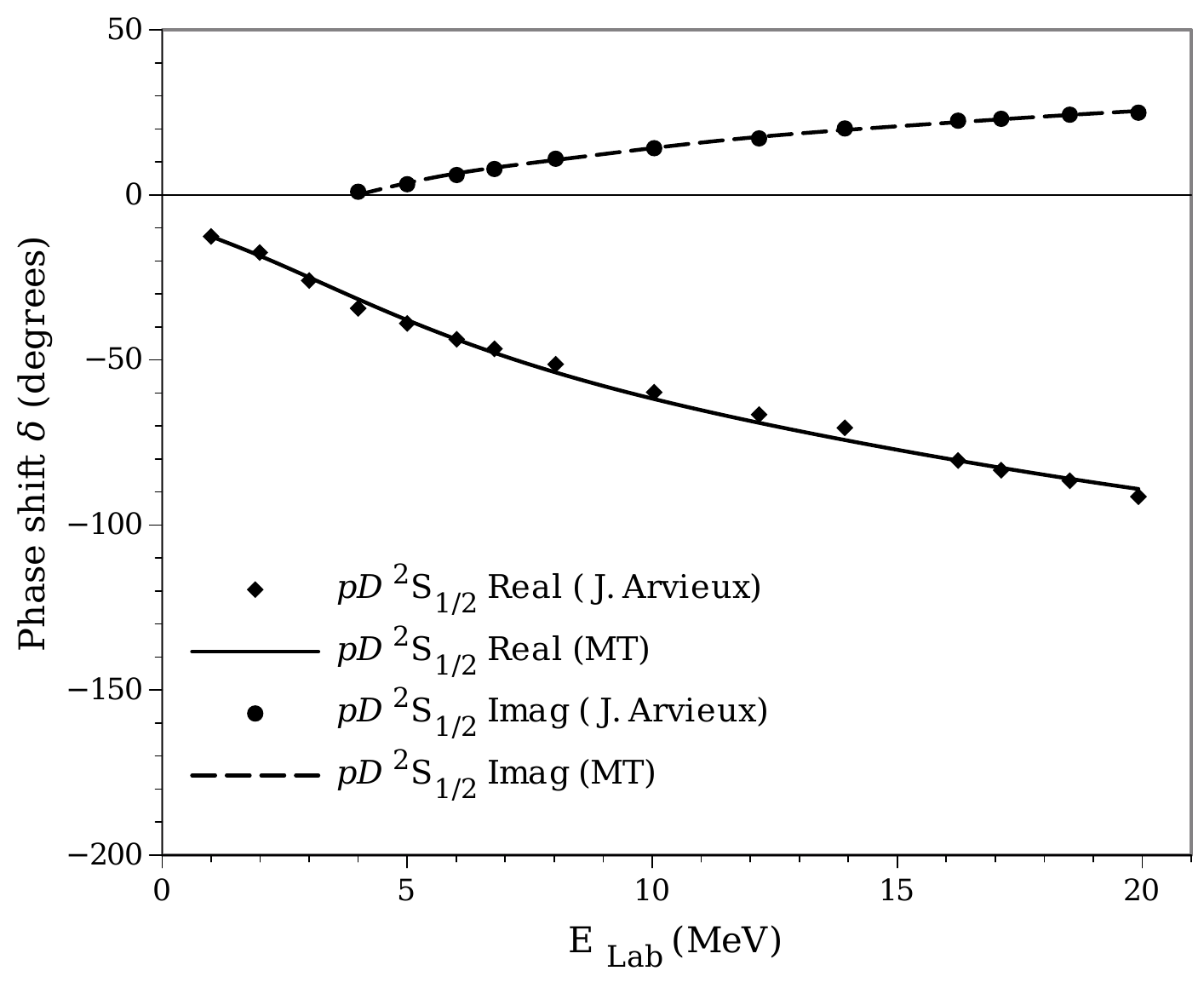}}
{\includegraphics[height = 7 cm, width =8.1cm]{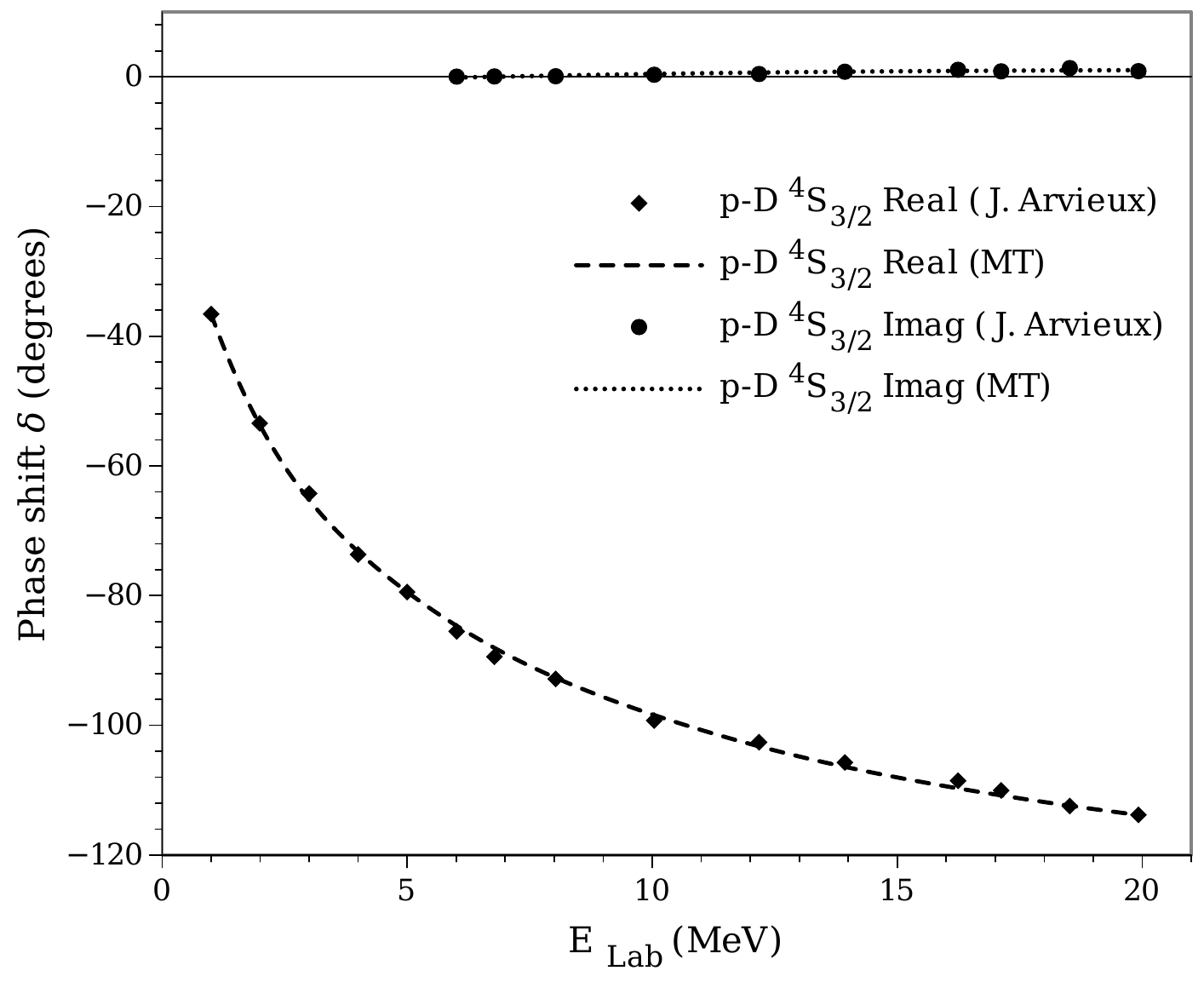}}
\caption{Real and imaginary parts of scattering phase shifts (SPS) for $^2S_{1/2}$ (\textit{left}) and $^4S_{3/2}$ (\textit{right}) states of $pd$ system, plotted against laboratory energy ranging between (1-19) $MeV$.}
\label{fig4}
\end{figure*}

The model parameters for $^2S_{1/2}$ and $^4S_{3/2}$ states of \textit{nd} and \textit{pd} systems are given in Table \ref{parameters}. The corresponding real and imaginary potentials are plotted in Fig. \ref{fig1}(a) and Fig. \ref{fig1}(b) respectively. It should be noted that for $^2S_{1/2}$ states of $nd$ and $pd$ systems, the real part of SPS is negative and therefore corresponding potential should be positive. But since $^2S_{1/2}$ state is the bound state of both $nd$ and $pd$ systems, therefore $180$\textdegree~is added to the SPS in order to match with the standard results of Huber \cite{Huber} and Arvieux \cite{Arvieux}.\\
\begin{table}[h!]
\centering
\caption{Model parameters ($V_{R}, V_{A}, \mu_{A} $) of MT potential and Mean absolute error (MAE), for real and imaginary parts of $^2S_{1/2}$ and $^4S_{3/2}$ states for (a) \textit{nD} and (b) \textit{pD}}
\begin{ruledtabular}
\begin{tabular}{l| c c | c c}
System & $^2S_{1/2}$ & MAE & $^4S_{3/2}$ & MAE\\
\hline
$nd$ (Real) & (22360.93, 4078.79, 1.16)&1.32  & \ \ (11.82, 3.55, 0.25)&0.19 \\
\hline
$nd$ (Imag) &  (114.11, 42.77, 0.65)&0.061& \ \ (490.85, 162.73, 2.94)& 0.06 \\
\hline \hline
$pd$ (Real) &  (913.67, 509.53, 0.56)&1.47 & \ \ (158.68, 37.66 , 0.14)& 0.55 \\
\hline
$pd$ (Imag) & (460.71, 80.19, 0.85)&0.62 & \ \ (1.27 $\times$ $10^{-13}$, 1.49, 0.058)&0.14 \\
\end{tabular}
\end{ruledtabular}
\label{parameters}
\end{table}
The effect of Coulomb potential is clearly visible in Fig. \ref{fig1}, since real potential corresponding to $^2S_{1/2}$ ground state of $nd$ system is deeper as compared to that for $pd$ system. The $^4S_{3/2}$ is not a bound state in both systems and, therefore corresponding real potentials are both positive. Yet one can observe that $pd$ system tends to create repulsion far earlier than $nd$ system.\\
The obtained real and imaginary parts of scattering phase shifts for $nd$ and $pd$ systems, have been plotted in Fig. \ref{fig3} and Fig. \ref{fig4} respectively. The standard data of Huber \cite{Huber} for $nd$ and J. Arvieux \cite{Arvieux} for $pd$ have also been shown in respective figures. 

The observed SPS for real and imaginary parts have been tabulated alongside the expected values in Table \ref{ndpd}. For $nd$ system, the obtained SPS are very close to that of standard data \cite{Huber} up-to 10 $MeV$, but for data points at 19 $MeV$, there is a discrepency of $5$\textdegree. In case of $pd$ system, the real as well as imaginary parts of SPS are matching well with standard data \cite{Arvieux} except few values which have about $3.2$\textdegree variance.

\begin{table*}[htp!]
\caption{Validation of obtained real and imaginary part of SPS, for doublet ($^2S_{1/2}$) and quartet ($^4S_{3/2}$) states of $nd$ and $pd$ scattering, w.r.t standard data \cite{Huber, Arvieux} respectively, for lab energies in range 1-19 MeV}
\begin{center}
\scalebox{0.89}{
\begin{ruledtabular}
\begin{tabular}{p{1.2cm}p{1.8cm}p{2cm}p{1.8cm}p{2cm}p{1.8cm}p{2cm}p{1.8cm}p{2.4cm}}
$E_{lab}$ &~ Real & ~Real & ~Imag & ~Imag & ~Real & ~Real & ~Imag & ~Imag \\
(MeV) & ~~\cite{Huber, Arvieux} & ~(MT) & ~ \cite{Huber, Arvieux} & ~(MT) &~ \cite{Huber, Arvieux} &~ (MT) &~ \cite{Huber, Arvieux} & (MT)\\
& \multicolumn{4}{c}{$^2S_{1/2}$} & \multicolumn{4}{c}{$^4S_{3/2}$}\\
\hline
\multicolumn{9}{c}{$nd$}\\
\hline	
1 & 163.9 &  \ \    163.9  &  --  &  --  & -0.987 & -0.987 & -- &--\\
3 &  147.9 & \ \  150.5   &  --    &  --  & -3.86 & -4.05 & -- &--\\
4.5 &  142.8  & \ \  142.8  &  2.2   &  2.2  & -5.29 & -5.29 & 0.0 & 0.0\\
6.5 & 134.2 & \ \  133.9    &  6.9  &  6.9  & -6.45 & -6.30 & 0.04 & 0.04 \\
10.5 &  119.2 & \ \  119.2  &  15.5  & 15.3 & -7.42 & -7.42 & 0.32 & 0.33 \\
19 &  90.2 & \ \  95.2   &  25.9  &  25.9 & -7.58 & -8.24 & 1.34 & 1.181 \\
\hline
\small{MAE}& &  \ \ 1.32 & & 0.06 && 0.19 & & 0.06 \\
\hline \hline
\multicolumn{9}{c}{$pd$}\\
\hline	
1 & 167.42 &   \ \   166.309  &  --  &  --   & -36.59 & -36.59 & -- & --\\
1.99 &  162.55 & \ \  161.169   &  --    &  --  & -53.43 & -53.68 & --& --\\
3 &  154.07  & \ \  155.221   &  --   &  --  & -64.26 & -65.19 & -- & --\\
4 & 145.65 & \ \   148.902    &  0.961    &  0.961  & -73.66 & -73.22 & --& -- \\
5 &  141.06 & \ \  142.736  &  3.242    & 2.977 & -79.45 & -79.45 & -- & --\\
6.01 &  136.25 & \ \  136.991   &  6.037    &  6.037  & -85.52 & -84.64 & 0.029 & -0.112 \\
6.78 & 133.38 &  \ \   132.925  &  7.862  &  8.095   & -89.42 & -88.06 & 0.057 & 0.006\\
8.03 &  128.69 & \ \  127.015   &  10.964   &  10.97  & -92.85 & -92.66 & 0.086& 0.186 \\
10.04 &  120.21 & \ \  118.864   &  14.161   &  15.125  & -99.27 & -98.42 & 0.317 & 0.442\\
12.18 & 113.45 & \ \   111.543    &  17.127  &  18.434  & -102.6 & -103.24 & 0.433 & 0.655\\
13.93 & 109.43 & \ \  106.207  &  20.145    & 20.145  & -105.74 & -106.40 & 0.784 & 0.787\\
16.24 &  99.52 & \ \  99.794   &  22.496    &  21.703  & -108.55 & -109.72 &  1.080  &  0.911\\
17.12 & 96.6 & \ \  97.516    &  23.067    &  22.193   & -110.04 & -110.81 & 0.843 & 0.946 \\
18.52 &  93.39 & \ \  94.072  &  24.311    & 22.853  & -112.45 & -112.40 &  1.349 &  0.989  \\
19.92 &  88.58 & \ \  90.846   &  24.92   &  23.34  & -113.82 & -113.82 & 0.873 & 1.021\\
\hline
\small{MAE}& &  \ \ 1.47 & & 0.62 && 0.55 & & 0.14 \\
\end{tabular}
\end{ruledtabular}}
\end{center}
\label{ndpd}
\end{table*}

\begin{figure*}[htp]
\centering
{\includegraphics[height = 6.5 cm, width =8cm]{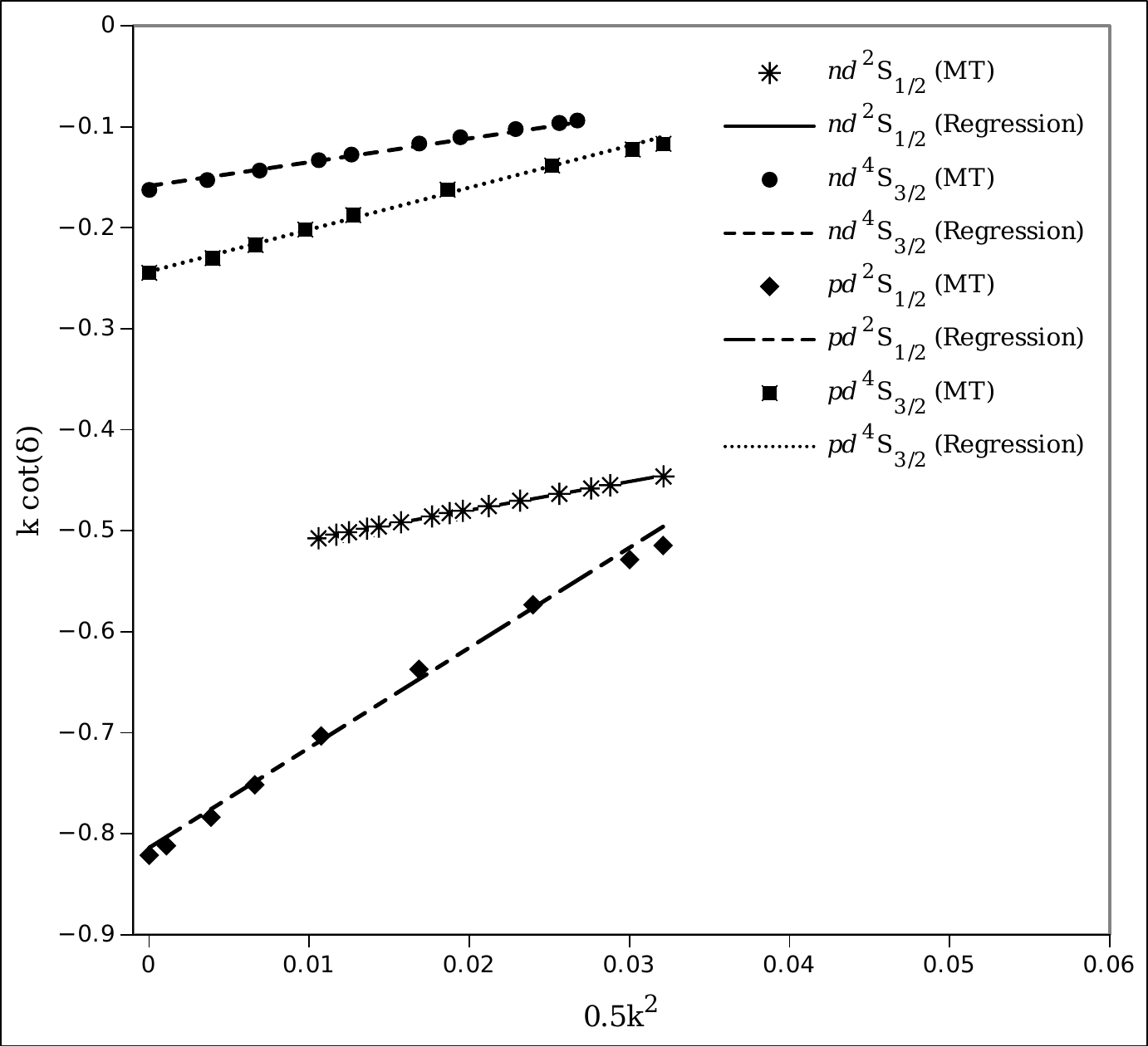}}
\caption{Plot of $(0.5 \times k^2)$ $vs$ $k cot(\delta)$ along with the regression lines for $^2S_{1/2}$ and $^4S_{3/2}$ states of \textit{nd} and \textit{pd} systems.}
\label{fig5}
\end{figure*}

In order to determine scattering parameters for doublet $^2S_{1/2}$ and quartet $^4S_{3/2}$ states of \textit{nd} and \textit{pd} scattering, only low energy data points are considered for lab energies up-to 3 $MeV$. At these low energies below breakup threshold, only real SPS are available. The $(0.5 \times k^2)$ $vs$ $k cot(\delta)$ plots for $^2S_{1/2}$ and $^4S_{3/2}$ states of \textit{nd} and \textit{pd} systems are shown in Fig. \ref{fig5}.
Using slope and intercept obtained from regression analysis, respective scattering parameters for all the channels are determined, and are presented in Table \ref{scat}.
\begin{table}[h]
\centering
\caption{Scattering parameters ($a (fm)$, $r_0 (fm)$) for $^2S_{1/2}$ and $^4S_{3/2}$ states of \textit{nd} and \textit{pd} scattering.}
\begin{tabular}{p{2.2cm}|p{2.9cm}p{2.9cm}}
\hline \hline
System & ~~~~$^2S_{1/2}$ & ~~~~$^4S_{3/2}$ \\
\hline
$nd$ & (1.86, 2.83) & (6.31, 2.35) \\
\hline
$pd$ & (1.54, 3.92) & (11.25, 1.56) \\
\hline \hline
\end{tabular}
\label{scat}
\end{table}
\clearpage
\section{Conclusion} 
Malfliet-Tjon potential has been considered as mathematical function to form both real and imaginary parts of an optical potential. By solving respective phase equations, numerically using RK-5 technique, for both real and imaginary parts of the optical potential, the respective real and imaginary scattering phase shifts for $nd$ and $pd$ systems have been obtained. The phase shifts are obtained for lab energies within elastic region of both $nd$ and $pd$ systems, ranging between 1 to 19 $MeV$. The mean absolute error (MAE) with respect to standard data of Huber $et.al.$ and J.Arvieux for all the states has been found to be less than 2. Also the scattering parameters i.e scattering length `$a$' and effective range `$r_0$' are found to be matching very well with experimental data.\\

\section{DISCLOSURE}
\begin{itemize}
    \item Funding: No funding was allocated for this work.
    \item Conflict of Interest: The authors declare that they have no conflict of interest.
\end{itemize}


\clearpage 

\end{document}